\documentclass{article}
\usepackage[utf8]{inputenc}
\usepackage{setspace}
\usepackage{indentfirst}
\usepackage{bbold}
\usepackage{amsmath}
\usepackage{amssymb}
\usepackage{mathtools}
\usepackage[round]{natbib}
\DeclarePairedDelimiter\ceil{\lceil}{\rceil}
\DeclarePairedDelimiter\floor{\lfloor}{\rfloor}

\begin{document}

\begin{titlepage}
\begin{center}
    \huge
    \vspace*{3cm}
    \textbf{Selecting Proportional Juries}\\
    \vspace{6cm}
    Eisho Takatsuji\footnote{Northwestern University, Department of Economics}\\
    \vspace{1cm}
    \LARGE
    June 10, 2026
\end{center}
\end{titlepage}



\newpage

\begin{center}
    \textbf{Abstract}\footnote{Many thanks to Bruno Strulovici, Annie Liang, Ezra Friedman, Stephanie Didwania, and Priyanka Goonetilleke for greatly helpful comments and support.}
\end{center}

I study the problem of selecting a jury from a qualified panel. Every jurisdiction in the United States (with the exception of Arizona) allows litigants to exercise peremptory strikes on the panel to eliminate potential jurors, with the remaining individuals being seated on the final jury. However, such a system prevents litigants from guaranteeing that the jury is proportional with respect to their preferred characteristic, particularly when only a minority of panel members exhibit that characteristic. Although some legal scholars have suggested affirmative selection, wherein litigants choose whom they want on the jury, as an alternative, this system also does not guarantee proportional juries. In this paper, I formalize ``proportionality" and show that there exists a family of methods that mix peremptory strikes and affirmative selection that allow litigants to guarantee that the jury is proportional with respect to their preferences over the panel.

\newpage

\section{Introduction}

Jury selection in the United States begins with a group of people, referred to as a venire, who are randomly chosen from the pool of jury-eligible citizens from the relevant jurisdiction. A process known as voir dire is then conducted on the venire so that individuals may be dismissed for cause (e.g. if they cannot be unbiased) or excused for hardship. In general, the judge decides which individuals to excuse in this stage (perhaps upon motions of the litigants), and there is no limit to the number of for-cause strikes or excusals that may happen. The individuals who remain after this stage make up the \textit{qualified panel}. Every jurisdiction in the United States, both federal and state—with the sole exception of Arizona—allows each party to then exercise peremptory challenges on members of the qualified panel to further strike prospective jurors from eligibility to be on the final jury. The individuals who remain after all peremptory challenges have been exercised make up the final seated jury.

The problem I am interested in is the selection of a seated jury $J$ of predetermined size $|J|$ from a qualified jury panel $P$ ($|J| \leq |P|$). In particular, my objective is to create a method that guarantees such selection is proportional: if a litigant likes some proportion of $P$, they should be able to guarantee that they like at least that proportion of $J$, lest one litigant get an ``unfair advantage" over the other wrt the distribution of $P$. For example, consider a panel of 100 people, of whom 10 must be chosen to make up the seated jury. Suppose that 70 of those 100 people are blue, and that 30 are red. Further suppose that litigant 1 prefers blue jurors and that litigant 2 prefers red jurors. Then, for a jury selection system to be proportional, it would need to be the case that under that system, 1) litigant 1 can guarantee that at least 7 of the seated jurors are blue and 2) litigant 2 can guarantee that at least 3 of the seated jurors are red.

The primary method in use today, peremptory strikes, does not lead to ``proportional" juries because it may result in the under-representation of minority characteristics in the presence of a litigant who exercises their strikes on said minority. For example, in the above example, let each litigant exercise 45 peremptory strikes. Then litigant 2 would not be able to guarantee that at least 3 seated jurors are red, because litigant 1 would be able to exercise peremptory strikes (of which they have 45) on every red individual (of which there are 30).

In light of this issue with peremptory strikes, several legal scholars have proposed the alternative system of affirmative selection: each litigant may affirmatively choose which individuals from the panel will be seated on the final jury. However, although affirmative selection does mitigate the issue of under-representation of minorities, it still does not lead to ``proportional" juries because it faces the converse issue of over-representation of minorities. Again considering the above example, suppose that each litigant receives 5 affirmative choices (``affirmatives") instead of 45 peremptory strikes. Then it would be litigant 1 who cannot guarantee that blue individuals are proportionally represented on the jury, because litigant 2 would be able to exercise all 5 of their affirmatives on red individuals.

Although neither a pure system of peremptory strikes nor a pure system of affirmative selection is proportional, few scholars have given much formal analysis to the class of what I call hybrid affirmative-peremptory methods (HAPs), wherein litigants interweave the exercise of peremptory strikes and affirmative selection. In this paper, I formalize the class of HAPs and show that there exists a subset of HAPs that are proportional.

At a high level, given a panel of size $|P|$ from which $|J|$ jurors must be chosen, an HAP is a $|P|$-step procedure that, at each step, designates one of the litigants to take a specified action (either peremptory or affirmative) on a potential juror in $P$ (who has not been acted upon in a prior step) of their choice. Such an HAP will be proportional if and only if, at every step, the ratio between (1) the number of affirmatives litigant 1 has exercised until that step and (2) the number of peremptories litigant 2 has exercised until that step (and vice versa) falls within certain bounds.

I also show that such methods are proportional not only with respect to litigants who have two-tiered preferences over the jury (i.e. ``like" vs ``dislike"), but also with respect to litigants who have arbitrarily multi-tiered preferences over the jury. For example, if $P$ comprises not only red and blue individuals, but also purple individuals, it may be that a litigant has a 3-tiered preference: blue over purple over red. In this case, I show that the family of proportional HAP methods would allow this litigant to guarantee (1) that the proportion of blue individuals in $J$ is at least the proportion of blue individuals in $P$ (rounded down); (2) that the proportion of blue-or-purple individuals in $J$ is at least the proportion of blue-or-purple individuals in $P$ (rounded down); and (3) that the proportion of blue-or-purple-or-red individuals in $J$ is at least the proportion of blue-or-purple-or-red individuals in $P$ (although that is trivial in this example).

The paper proceeds as follows. Section 2 is a brief overview of related literature. Section 3 provides a formal definition of the ``proportionality" of a jury selection method, both for two-tiered and for multi-tiered preferences. Sections 4 and 5 formalize the intuition that neither purely peremptory strike based methods nor purely affirmative selection based methods are proportional. Section 6 defines HAP methods and contains the main theorems of the paper: the necessary and sufficient conditions for an HAP method to be proportional and the existence of such methods. Section 7 considers an even stronger notion of proportionality that is inspired by the Droop quota from analysis of electoral systems. Section 8 provides a brief proof that any ``proportional" jury selection method will allow each litigant to select a better-than-approximately- ``representative" jury. Section 9 discusses some subtleties relating to the practical implementation of HAP methods. Section 10 discusses impossibility results when generalizing to settings with more than two litigants.

\section{Related Literature}
Although a number of legal scholars have raised the idea of using jury selection methods that allow litigants to mix the exercise of peremptories with that of affirmatives (e.g. \cite{altman1986affirmative}, \cite{zeisel1987affirmative}, \cite{coke1994lady}), this paper is the first to formalize the class of hybrid affirmative-peremptory (HAP) jury selection methods, as well as the first to examine the conditions under which such methods are ``proportional."

To the author's knowledge, there is one working paper that studies the setting of jury selection in a manner similar to this paper: \cite{cohen2026adversarial} use a method inspired by the cut-and-choose method from the cake-cutting literature to devise a method that selects a jury that is maximally representative of the population with respect to three ``representativeness" statistics.\footnote{The author would like to note that his work on the instant paper was done independently of and contemporaneously with Cohen et al.'s work.}

There have also been previous theoretical papers that specifically study various implementations of purely peremptory challenge based systems. \cite{brams1978optimal} provide a game theoretic analysis of jury selection under a strike-and-replace jury selection system, wherein panel members are presented to, and may be struck by, each litigant one-by-one. \cite{moro2024exclusion} compare the performance of the strike-and-replace system and that of the struck system (under which the whole panel is first presented to the litigants, who may then exercise their strikes on any members of the panel) in selecting (1) jurors from minority groups and (2) extreme jurors. \cite{flanagan2015peremptory} takes a step back and shows that any purely peremptory challenge based system may be more likely to result in homogeneous-and-extreme juries than random selection is, lending credence to the notion that a system of peremptory strikes does not seat proportional juries.

\section{Formalization of proportionality}

Consider a litigant who, given a panel $P$, wants to ensure the representation of some subset of the panel $S \subseteq P$ on the final jury $J$. For example, $S$ could represent the subset of prospective jurors who are black, who are white, who are women, who are men, who have children, who listen to a certain news outlet, etc. A litigant may choose any characteristic for which it wishes to ensure representation.

Given a panel $P$ and a subset $S$, a jury $J$ is deemed ``$(P,S)$-proportional" if 
\begin{equation}
    \forall q \in \mathbb{N} \text{ s.t. } |S| \geq \frac{|P|}{|J|}q , |S \cap J| \geq q
\end{equation}

Note that condition (1) is equivalent to the following: $|S \cap J| \geq \floor*{\frac{|J|}{|P|} {|S|}}$.

Intuitively, if $S$ makes up at least as large a proportion of $P$ as a set of cardinality $q$ does of $J$, then at least $q$ members of $S$ should be chosen to sit on the final jury. For example, if the qualified panel $P$ is 30 people, out of which a jury $J$ of 12 people must be chosen, and the set $S \subseteq P$ has 10 members, then the final jury $J$ will be $(P,S)$-proportional if it contains at least 4 members of $S$, because $\frac{10}{30} \geq \frac{4}{12}$, but $\frac{10}{30} \not\geq \frac{5}{12}$.

A jury selection method will be deemed ``quasi-$(|P|,|J|)$-proportional wrt litigant $i$" if $\forall P$ of cardinality $|P|$, and $S \subseteq P$, $i$ can guarantee that the seated jury $J$ (of size $|J|$) will be $(P,S)$-proportional. (A litigant can ``guarantee" a result, given a jury selection method, if they have a strategy such that, for any strategy their opponent employs, the result will hold). In other words, this property is true if a litigant can guarantee that no matter the subset $S$ of the panel $P$ for which they want to ensure representation, the members of $S$ retain a proportionate presence in the seated jury $J$.

A jury selection method will be deemed ``quasi-$(|P|,|J|)$-proportional" if it is quasi-$(|P|,|J|)$-proportional wrt both litigants.

Intuitively, quasi-$(|P|,|J|)$-proportionality captures the notion that each litigant should, given a two-tiered preference, be able to guarantee the proportionality of $J$ with respect to their preferred subset of $P$. The following definitions capture the more general case of arbitrarily multi-tiered preferences.

Given a panel $P$ and a sequence of nested subsets $\textbf{S} = \{S_i \subseteq P\}_{i=1}^I$ such that $S_i \subset S_{i+1}$ for all $i < I$, the selected jury $J$ will be deemed ``$(P,\textbf{S})$-proportional" if $\forall S_i \in \textbf{S}$, $J$ is $(P,S_i)$-proportional.

A jury selection method will be deemed ``$(|P|,|J|)$-proportional wrt litigant $i$" if $\forall P$ of cardinality $|P|$, and sequence of nested subsets $\textbf{S}$, $i$ can guarantee that the seated jury $J$ (of size $|J|$) is $(P,\textbf{S})$-proportional.

A jury selection method will be deemed ``$(|P|,|J|)$-proportional" if it is $(|P|,|J|)$-proportional wrt both litigants.

Intuitively, $(|P|,|J|)$-proportionality captures the notion that each litigant should, given an arbitrarily multi-tiered preference, be able to guarantee the proportionality of $J$ with respect to every subset of $P$ for which they prefer every element of that subset to every non-element of that subset.

Note that in both senses of the terms, $(|P|,|J|)$-proportionality implies quasi-$(|P|,|J|)$-proportionality.

\section{Peremptory strikes are not quasi-proportional.}
A system of peremptory challenges is not a quasi-$(|P|,|J|)$-proportional jury selection method whenever $|J|>2, |P| \geq \frac{|J|^2}{|J|-2}$ (note, for example, that this is true whenever $4 \leq |J| \leq |P|-4$). I first formally define what a system of peremptory challenges is.

Given that $|J|$ jurors must be chosen from a panel of size $|P|$ to make up the jury $J$, the peremptory method is defined by an ordered pair $(b_1,b_2)$ such that $b_1 + b_2 = |P| - |J|$, where $b_1$ is the number of peremptory strikes that litigant 1 may exercise and $b_2$ is the number of peremptory strikes litigant 2 may exercise.

Consider arbitrary $P$ and $|J|$ such that $|P| \geq \frac{|J|^2}{|J|-2}$. Further consider a subset $S \subseteq P$ such that $|S| = \ceil*{\frac{|P|-|J|}{2}}$. For $J$ to be $(P,S)$-proportional, it must be that $|S \cap J| \geq 1$, because $\floor*{\frac{|J|}{|P|} |S|} = \floor*{\frac{|J|}{|P|} \ceil*{\frac{|P|-|J|}{2}}} \geq \floor*{\frac{|J|}{|P|} \frac{|P|-|J|}{2}} \geq 1$. This, in turn, is because:

\begin{align*}
    \frac{|J|}{|P|} \frac{|P|-|J|}{2} \geq 1 \iff |J||P|-|J|^2 \geq 2|P| \iff |P| \geq \frac{|J|^2}{|J|-2}
\end{align*}

However, because a total of $|P|-|J|$ peremptory strikes must be exercised by the two litigants, one of the litigants must have at least $\ceil*{\frac{|P|-|J|}{2}}$ peremptory strikes. Thus, consider a litigant for whom the opposing litigant has at least $\ceil*{\frac{|P|-|J|}{2}}$ peremptory strikes. This jury selection method is not quasi-$(|P|,|J|)$-proportional wrt that litigant because no matter what strategy that litigant chooses, the opposing litigant can simply strike all $\ceil*{\frac{|P|-|J|}{2}}$ members of $S$ from being seated on the final jury, so that $|S \cap J| = 0$.

\section{Affirmative selection is not quasi-proportional.}
A system of affirmative selection is not a quasi-$(|P|,|J|)$-proportional jury selection method whenever $|J|>2, |P| \geq \frac{|J|^2}{|J|-2}$ (note, for example, that this is true whenever $4 \leq |J| \leq |P|-4$). I first formally define what a system of affirmative selection is.

Given that $|J|$ jurors must be chosen from a panel of size $|P|$ to make up the jury $J$, the affirmative selection method is defined by an ordered pair $(a_1,a_2)$ such that $a_1 + a_2 = |J|$, where $a_1$ is the number of affirmatives that litigant 1 may exercise (i.e. the number of jurors litigant 1 may affirmatively choose) and $a_2$ is the number of affirmatives litigant 2 may exercise.

Consider arbitrary $P$ and $|J|$ such that $|P| \geq \frac{|J|^2}{|J|-2}$. Further consider a subset $S \subseteq P$ such that $|S| = |P|-\ceil*{\frac{|J|}{2}}$. For $J$ to be $(P,S)$-proportional, it must be that $|S \cap J| > \floor*{\frac{|J|}{2}}$, because $\floor*{\frac{|J|}{|P|} |S|} = \floor*{\frac{|J|}{|P|} (|P|-\ceil*{\frac{|J|}{2}})} =  |J|-\ceil*{\frac{|J|}{|P|}\ceil*{\frac{|J|}{2}}} > |J| - \ceil*{\frac{|J|}{2}} = \floor*{\frac{|J|}{2}}$. This, in turn, is because:

\begin{align*}
     \ceil*{\frac{|J|}{|P|}\ceil*{\frac{|J|}{2}}} < \ceil*{\frac{|J|}{2}} & \iff \frac{|J|}{|P|} \leq \frac{\ceil*{\frac{|J|}{2}}-1}{\ceil*{\frac{|J|}{2}}}\\
     &\impliedby \frac{|J|}{|P|} \leq \frac{\frac{|J|}{2}-1}{\frac{|J|}{2}}\\
     &\iff |P| \geq \frac{|J|^2}{|J|-2}
\end{align*}

The $\impliedby$ is true because $\frac{x-1}{x}$ is increasing in $x$ and $\ceil{x} \geq x$.

However, because a total of $|J|$ affirmatives must be exercised by the two litigants, one of the litigants must have at least $\ceil*{\frac{|J|}{2}}$ affirmatives. Thus, consider a litigant for whom the opposing litigant has at least $\ceil*{\frac{|J|}{2}}$ affirmatives. This jury selection method is not quasi-$(|P|,|J|)$-proportional wrt that litigant because no matter what strategy that litigant chooses, the opposing litigant can simply choose all $\ceil*{\frac{|J|}{2}}$ members of $S^C$ to be seated on the final jury, so that $|S \cap J| \leq |J| - \ceil*{\frac{|J|}{2}} = \floor*{\frac{|J|}{2}}$.

\section{There exist hybrid affirmative-peremptory methods that are proportional.}

Although neither the pure peremptory strike method nor the pure affirmative selection method is, in general, quasi-proportional, this does not mean that there is no way to combine the two into a hybrid method that \textit{is} always quasi-proportional.

I first formally define the family of hybrid affirmative-peremptory (HAP) systems. As before, let there be a qualified panel of size $|P|$ from which a jury $J$ of cardinality $|J| \leq |P|$ must be chosen. Denote the two litigants L1 and L2.

A $(|P|,|J|)$-HAP method is defined by a sequence of ordered quadruples $\{(a_{1,t}, a_{2,t}, b_{1,t}, b_{2,t})\}_{t=0}^{|P|}$. $a_{1,t}$ denotes the number of affirmatives that L1 has had the opportunity to exercise until time $t$. (Note: $a_{1,|P|}$ denotes the \textit{total} number of affirmatives L1 gets during the entirety of jury selection). $a_{2,t}$ is the number of affirmatives that L2 has had until time $t$ (Note: $a_{2,|P|}$ denotes the total number of affirmatives L2 gets). $b_{1,t}$ is the number of peremptories that L1 has had until time $t$ (Note: $b_{1,|P|}$ denotes the total number of peremptories L1 gets). $b_{2,t}$ is the number of peremptories that L2 has had until time $t$ (Note: $b_{2,|P|}$ denotes the total number of peremptories L2 gets).

The initial conditions are: $a_{1,0} = a_{2,0} = b_{1,0} = b_{2,0} = 0$. After that, $\forall 1 \leq t \leq |P|$, exactly one of the four values in the ordered quadruple increases by 1 from the previous time ($t-1$). That is, at each time, exactly 1 of the following happens:
\begin{itemize}
    \item $a_{1,t} = a_{1,t-1} + 1$ (i.e. L1 may exercise 1 affirmative at $t$)
    \item $a_{2,t} = a_{2,t-1} + 1$ (i.e. L2 may exercise 1 affirmative at $t$)
    \item $b_{1,t} = b_{1,t-1} + 1$ (i.e. L1 may exercise 1 peremptory at $t$)
    \item $b_{2,t} = b_{2,t-1} + 1$ (i.e. L2 may exercise 1 peremptory at $t$)
\end{itemize}

In total, $|P|$ actions are taken. This represents that the sum of the total number of peremptories and the total number of affirmatives must be the total number of people on the qualified panel. This is why $t$ ends at $|P|$.

In addition, it must be that $a_{1,|P|} + a_{2,|P|} = |J|$ and that $b_{1,|P|} + b_{2,|P|} = |P| - |J|$. These equations represent that (1) the total number of affirmatives exercised should equal the size of the jury and (2) the total number of peremptories exercised should equal the excess of the size of the qualified panel over the size of the jury.

Define a given individual $p \in P$ to be ``available at time $t$" if no peremptory or affirmative has been exercised on them until time $t$. For all $t$, litigants may only exercise an affirmative or peremptory on individuals who are available at $t$. Intuitively, no panel member who has been acted on (whether struck or chosen) may be acted on again.

Note that the pure peremptory system is just one type of the generalized hybrid method, wherein only $b_1$ and $b_2$ increase (and $a_1$ and $a_2$ remain at 0) until $t = |P| - |J|$. Similarly, the pure affirmative selection system is just one type of the generalized hybrid method, wherein only $a_1$ and $a_2$ increase (and $b_1$ and $b_2$ remain at 0) until $t = |J|$.\\

\textbf{NOTE:} For convenience, when discussing a $(|P|,|J|)$-HAP method, I will omit the ``$(|P|,|J|)$-" when referring to any proportionality conditions.\\

\textit{Theorem 1:} Any $(|P|,|J|)$-HAP method that has the following properties is quasi-proportional.

\begin{align}
    \forall i,j \in \{1,2\} (i \neq j), \forall t: \frac{|P|-|J|}{|J|}a_{j,t} - \frac{|P|}{|J|} < b_{i,t} < \frac{|P|-|J|}{|J|}a_{j,t} + \frac{|P|}{|J|}
\end{align}

\textit{Proof:} WLOG, I will prove that a $(|P|,|J|)$-HAP method is quasi-proportional wrt L1 \textit{if}

\begin{align}
    \forall t, b_{1,t} > \frac{|P|-|J|}{|J|}a_{2,t} - \frac{|P|}{|J|} \\    
    \forall t, b_{2,t} < \frac{|P|-|J|}{|J|}a_{1,t} + \frac{|P|}{|J|}
\end{align}

Combining this with the opposite inequalities for L2 yields the above result.\\

Consider an arbitrary panel $P$ of size $|P|$ and subset $S \subseteq P$, where a jury $J\subseteq P$ of size $|J|$ must be chosen. I will prove that if L1 adopts the following strategy (which I will call the ``ordering strategy wrt $S$"), then $J$ will be $(P,S)$-proportional regardless of L2's strategy:

\begin{itemize}
    \item L1 orders the individuals in $P$ into a sequence $\{p_i\}_{i=1}^{|P|}$, such that for $1 \leq i \leq |S|$, $p_i \in S$. That is, L1 orders the members of $S$ first.
    \item Whenever L1 has an affirmative, use it on the \textit{first} available individual, i.e., the available individual with the smallest $i$.
    \item Whenever L1 has a peremptory, use it on the \textit{last} available individual, i.e., the available individual with the largest $i$.
\end{itemize}

\textit{Lemma 1.1:} If $\exists t$ and $x,y \in \mathbb{N}$ s.t. $a_{1,t} = x$ and $b_{2,t} \leq y$, then $\forall S\subseteq P$ for which $|S| \geq x+y$, if L1 uses the ordering strategy wrt $S$, then $|S \cap J| \geq x$. In other words, if there is any time $t$ at which litigant 1 has $x$ affirmatives while litigant 2 has at most $y$ peremptories, then for any group $S$ which contains at least $x+y$ individuals, L1 can guarantee that at least $x$ of them end up on the final jury by employing the ordering strategy wrt $S$.\\

\textit{Proof:} I first introduce some notation. Given a set $S$, denote $S_{a,t}$ to be the subset of $S$ that is available (i.e. has not yet been acted on by either litigant) at the start of time $t$; and $J_t$ to be the subset of $P$ that have been affirmatively chosen until time $t$ (i.e. the members of the jury that have been locked in at $t$).

By assumption, $\exists t$ and $x,y \in \mathbb{N}$ s.t. $a_{1,t} = x$ and $b_{2,t} \leq y$. Call this time $T$. We want to show that  $\forall S\subseteq P$ for which $|S| \geq x+y$, if L1 uses the ordering strategy wrt $S$, then $|S \cap J| \geq x$. I will do this by using induction to show that until (and including) the first time at which $S_{a,t} = \emptyset$ (which must occur, because every member of $P$ is eventually acted upon), the following inequality is true: $|S \cap J_t| + \min\{a_{1,T} - a_{1,t}, |S_{a,t}|-(b_{2,T}-b_{2,t})\} \geq x$.

The inequality is true at $t=0$: $J_0 = \emptyset$, $S_{a,0} = S$, $a_{1,0} = 0$, and $b_{2,0} = 0$, so $|S \cap J_0| + \min\{a_{1,T} - a_{1,0}, |S_{a,0}|-(b_{2,T}-b_{2,0})\} = 0 + \min\{a_{1,T}, |S| - b_{2,T}\} = x$. (Recall that $|S| \geq x + y \geq a_{1,T} + b_{2,T}$).

If the inequality is true at the start of $t$ and $|S_{a,t}| > 0$, it will be true at the start of $t+1$, regardless of which litigant is assigned to take which action at $t$:

\begin{itemize}
    \item If L1 uses an affirmative, it will be on a member of $S_{a,t}$, due to the definition of the ordering strategy wrt $S$ and the assumption that $|S_{a,t}| > 0$. Thus, $|S \cap J_t|$ increases by 1. Also, $a_{1,t}$ increases by 1 and $|S_{a,t}|$ decreases by 1. So the value of the LHS of the inequality remains the same, and the inequality holds.

    \item If L1 uses a peremptory, it will either be on a member of $S_{a,t}$ or on a member of $S^C_{a,t}$ (the members of $S^C$ that are available at the start of $t$).
    \begin{itemize}
        \item If on a member of $S_{a,t}$, then $|S_{a,t}|$ decreases by 1 and no other value in the LHS changes. But if L1 is using a peremptory on a member of $S_{a,t}$ while using the ordering strategy wrt $S$, it must be that $P_{a,t} = S_{a,t}$ (that is, only members of $S$ are still available). In addition, if L1 is using a peremptory, it must be that $|P_{a,t}| > (b_{2,|P|}-b_{2,t}) + (a_{1,|P|} - a_{1,t}) \geq (b_{2,T}-b_{2,t}) + (a_{1,T} - a_{1,t})$. That is, there must be more available members of $P$ than the combined total of the number of L1's affirmatives and L2's peremptories. Thus, $|S_{a,t}|-(b_{2,T}-b_{2,t}) > a_{1,T} - a_{1,t}$, and the value of the $\min$ term would be unchanged overall. So the value of the LHS remains the same, and the inequality holds.

        \item If on a member of $S^C_{a,t}$, then nothing in the LHS changes, and the inequality holds.
    \end{itemize}

    \item If L2 uses an affirmative on a member of:
    \begin{itemize}
        \item $S_{a,t}$, then $|S \cap J_t|$ increases by 1 and $S_{a,t}$ decreases by 1. So the value of the LHS weakly increases, and the inequality holds.

        \item $S^C_{a,t}$, then nothing in the LHS changes, and the inequality holds.
    \end{itemize}

    \item If L2 uses a peremptory on a member of:
    \begin{itemize}
        \item $S_{a,t}$, then $|S_{a,t}|$ decreases by 1 and $b_{2,t}$ increases by 1, so the value of the LHS remains the same, and the inequality holds.

        \item $S^C_{a,t}$, then $b_{2,t}$ increases by 1, so the value of the LHS weakly increases, and the inequality holds.
    \end{itemize}

\end{itemize}

I have shown that $|S \cap J_t| + \min\{a_{1,T} - a_{1,t}, |S_{a,t}|-(b_{2,T}-b_{2,t})\} \geq x$ must hold until the start of the first $t$ at which $S_{a,t}=0$. Call this time $T'$. Note two things. First, $S \cap J_t$ cannot change after $T'$, because there are no available members of $S$ to be chosen to sit on $J$. Second, either $a_{1,T} - a_{1,T'}$ or $|S_{a,T'}|-(b_{2,T}-b_{2,T'}) = b_{2,T'} - b_{2,T}$ must be non-positive. This is because either $T' \geq T$, in which case $a_{1,T'} \geq a_{1,T}$, or $T' \leq T$, in which case $b_{2,T'} \leq b_{2,T}$. Thus, it must be that $|S \cap J| = |S \cap J_{T'}| \geq x$.

\hfill

\textit{Lemma 1.2:} Assuming eq. (4) is true, then $\forall r \leq a_{1,|P|}, \exists t$ s.t. $a_{1,t} = r$ and $b_{2,t} \leq \ceil*{\frac{|P|-|J|}{|J|}r}$. That is, for any number no greater than the total number of affirmatives that L1 gets, there will be a time at which L1 has that many affirmatives \textit{and} L2 has no greater than a threshold number of peremptories.\\

\textit{Proof:} First note that this is trivially true for $r=0$, because the inequality would be satisfied at $t=0$ (because $b_{2,0} = 0$).

For $r>0$, I prove this lemma by contrapositive, by showing that if it doesn't hold, then eq. (4) can't be true.

Suppose $\exists \text{ } 1 \leq r \leq a_{1,|P|}$ s.t. $\forall t$ for which $a_{1,t} = r$, $b_{2,t} > \ceil*{\frac{|P|-|J|}{|J|}a_{1,t}}$. Because $b_{2,t}$ and $\ceil*{\frac{|P|-|J|}{|J|}a_{1,t}}$ are always integers, this would mean that $b_{2,t} \geq \ceil*{\frac{|P|-|J|}{|J|}a_{1,t}} + 1$. In addition, because this must hold for all $t$ for which $a_{1,t} = r$, it must hold in particular for the $t'$ at which $a_{1,t'-1} = a_{1,t'}-1 = r-1$ (that is, for the first $t'$ at which L1 had $r$ affirmatives). Define $t''\equiv t'-1$ (the last $t$ at which L1 had $r-1$ affirmatives). Note that because only one action can happen at each time, it must be that $b_{2,t''} = b_{2,t'}$ (that is, L2 still had the same number of peremptories at $t''$ as she did at $t'$).

In this case, 
\begin{align*}
    \frac{|P|-|J|}{|J|}a_{1,t''} + \frac{|P|}{|J|} &= \frac{|P|-|J|}{|J|}(a_{1,t'}-1) + \frac{|P|}{|J|}\\
    &=\frac{|P|-|J|}{|J|}a_{1,t'} + 1\\
    &\leq \ceil*{\frac{|P|-|J|}{|J|}a_{1,t}} + 1\\
    &\leq b_{2,t'} = b_{2,t''}
\end{align*}

Note that this inequality \textit{violates} eq. (4). Thus, by contrapositive, if eq. (4) is true, then $\forall r \leq a_{1,|P|}, \exists t$ s.t. $a_{1,t} = r$ and $b_{2,t} \leq \ceil*{\frac{|P|-|J|}{|J|}r}$.\\

\textit{Corollary 1.1:} If eq. (4) is true, then Lemmas 1.1 and 1.2 together imply that $\forall 0 \leq r \leq a_{1,|P|}, \forall S \subseteq P$ for which $|S| \geq \ceil*{\frac{|P|}{|J|}r}$, L1 can guarantee that $|S \cap J| \geq r$ by using the ordering strategy wrt $S$. Note that $r + \ceil*{\frac{|P|-|J|}{|J|}r} = \ceil*{\frac{|P|}{|J|}r}$.\\

\textit{Lemma 1.3:} If $\exists t$ and $x,y \in \mathbb{N}$ s.t. $a_{2,t} = x$ and $b_{1,t} \geq y$, then $\forall S\subseteq P$ for which $|S^C| \leq x+y$, if L1 uses the ordering strategy wrt $S$, then $|S \cap J| \geq |J| - x$ (equivalently, $|S^C \cap J| \leq x$). In other words, if there is any time $t$ at which L2 has $x$ affirmatives while L1 has at least $y$ peremptories, then for any group $S^C$ which contains at most $x+y$ individuals, L1 guarantees that at most $x$ of them end up on the final jury by employing the ordering strategy wrt $S$.\\

\textit{Proof:} By assumption, $\exists t$ and $x,y \in \mathbb{N}$ s.t. $a_{2,t} = x$ and $b_{1,t} \geq y$. Call this time $T$. We want to show that  $\forall S\subseteq P$ for which $|S^C| \leq x+y$, if L1 uses the ordering strategy wrt $S$, then $|S^C \cap J| \leq x$. I will do this by using induction to show that until (and including) the first time at which $S^C_{a,t} = \emptyset$, the following inequality is true: $|S^C \cap J_t| + \max\{a_{2,T} - a_{2,t}, |S^C_{a,t}|-(b_{1,T}-b_{1,t})\} \leq x$.

The inequality is true at $t=0$: $J_0 = \emptyset$, $S^C_{a,0} = S^C$, $a_{2,0} = 0$, and $b_{1,0} = 0$, so $|S^C \cap J_0| + \max\{a_{2,T} - a_{2,0}, |S^C_{a,0}|-(b_{1,T}-b_{1,0})\} = 0 + \max\{a_{2,T}, |S^C| - b_{1,T}\} = x$. (Recall that $|S^C| \leq x + y \leq a_{2,T} + b_{1,T}$).

If the inequality is true at the start of $t$ and $|S^C_{a,t}| > 0$, it will be true at the start of $t+1$, regardless of which litigant is assigned to take which action at $t$:

\begin{itemize}
    \item If L1 uses an affirmative on a member of:
    \begin{itemize}
        \item $S_{a,t}$, then nothing in the LHS changes and the inequality holds.
        
        \item $S^C_{a,t}$, then $|S^C \cap J_t|$ increases by 1 and $|S^C_{a,t}|$ decreases by 1. In addition, it must be that $|S^C_{a,t}|-(b_{1,T}-b_{1,t}) > a_{2,T} - a_{2,t}$. This is because (1) if L1 is using an affirmative on a member of $S^C_{a,t}$ while using the ordering strategy wrt $S$, it must be that $P_{a,t} = S^C_{a,t}$; and (2) if L1 is using an affirmative, it must be that $|P_{a,t}| > (b_{1,|P|}-b_{1,t}) + (a_{2,|P|} - a_{2,t}) \geq (b_{1,T}-b_{1,t}) + (a_{2,T} - a_{2,t})$. So the value of the LHS remains the same, and the inequality holds.
    \end{itemize}

    \item If L1 uses a peremptory, it will be on a member of $S^C_{a,t}$, due to the definition of the ordering strategy wrt $S$ and the assumption that $|S^C_{a,t}| > 0$. In this case, $|S^C_{a,t}|$ decreases by 1 and $b_{1,t}$ increases by 1, so the value of the LHS remains the same, and the inequality holds.

    \item If L2 uses an affirmative on a member of:
    \begin{itemize}
        \item $S_{a,t}$, then $a_{2,t}$ increases by 1. So the value of the LHS weakly decreases, and the inequality holds.

        \item $S^C_{a,t}$, then $|S^C \cap J_t|$ increases by 1, $a_{2,t}$ increases by 1 and $|S^C_{a,t}|$ decreases by 1. So the value of the LHS remains the same, and the inequality holds.
    \end{itemize}

    \item If L2 uses a peremptory on a member of:
    \begin{itemize}
        \item $S_{a,t}$, then nothing in the LHS changes, and the inequality holds.

        \item $S^C_{a,t}$, then $|S^C_{a,t}|$ decreases by 1. So the value of the LHS weakly decreases, and the inequality holds.
    \end{itemize}

\end{itemize}

I have shown that $|S^C \cap J_t| + \max\{a_{2,T} - a_{2,t}, |S^C_{a,t}|-(b_{1,T}-b_{1,t})\} \leq x$ must hold until the start of the first $t$ at which $S^C_{a,t}=0$. Call this time $T'$. Note two things. First, $S^C \cap J_t$ cannot change after $T'$, because there are no available members of $S^C$ to be chosen to sit on $J$. Second, either $a_{2,T} - a_{2,T'}$ or $|S^C_{a,T'}|-(b_{1,T}-b_{1,T'}) = b_{1,T'} - b_{1,T}$ must be non-negative. This is because either $T' \geq T$, in which case $b_{1,T'} \geq b_{1,T}$, or $T' \leq T$, in which case $a_{2,T'} \leq a_{2,T}$. Thus, it must be that $|S^C \cap J| = |S^C \cap J_{T'}| \leq x$.

\hfill

\textit{Lemma 1.4:} Assuming eq. (3) is true, then $\forall r < a_{2,|P|}, \exists t$ s.t. $a_{2,t} = r$ and $b_{1,t} \geq \floor*{\frac{|P|-|J|}{|J|}r}$. That is, for any number less than the total number of affirmatives that L2 gets, there will be a time at which L2 has that many affirmatives \textit{and} L1 has no fewer than a threshold number of peremptories.\\

\textit{Proof:} For $0\leq r<a_{2,|P|}$, I prove this lemma by contrapositive, by showing that if it doesn't hold, then eq. (3) can't be true.

Suppose $\exists \text{ } 0 \leq r < a_{2,|P|}$ s.t. $\forall t$ for which $a_{2,t} = r$, $b_{1,t} < \floor*{\frac{|P|-|J|}{|J|}a_{2,t}}$. Because $b_{1,t}$ and $\floor*{\frac{|P|-|J|}{|J|}a_{2,t}}$ are always integers, this would mean that $b_{1,t} \leq \floor*{\frac{|P|-|J|}{|J|}a_{2,t}} - 1$. In addition, because this must hold for all $t$ for which $a_{2,t} = r$, it must hold in particular for the $t'$ at which $a_{2,t'+1} = a_{2,t'}+1 = r+1$ (that is, for the last $t'$ at which L2 had $r$ affirmatives). Define $t''\equiv t'+1$ (the first $t$ at which L2 had $(r+1)$ affirmatives). Note that because only one action can happen at each time, it must be that $b_{1,t''} = b_{1,t'}$ (that is, L1 still had the same number of peremptories at $t'$ as he did at $t''$).

In this case, 
\begin{align*}
    \frac{|P|-|J|}{|J|}a_{2,t''} - \frac{|P|}{|J|} &= \frac{|P|-|J|}{|J|}(a_{2,t'}+1) - \frac{|P|}{|J|}\\
    &=\frac{|P|-|J|}{|J|}a_{2,t'} - 1\\
    &\geq \floor*{\frac{|P|-|J|}{|J|}a_{2,t}} -1\\
    &\geq b_{1,t'} = b_{1,t''}
\end{align*}

Note that this inequality \textit{violates} eq. (3). Thus, by contrapositive, if eq. (3) is true, then $\forall r < a_{2,|P|}, \exists t$ s.t. $a_{2,t} = r$ and $b_{1,t} \geq \floor*{\frac{|P|-|J|}{|J|}r}$.\\

\textit{Corollary 1.2:} If eq. (3) is true, then $\forall a_{1,|P|} < r \leq |J|, \forall S \subseteq P$ for which $|S| \geq \ceil*{\frac{|P|}{|J|}r}$, if L1 uses the ordering strategy wrt $S$, then $|S \cap J| \geq r$

\textit{Proof:}
\begin{itemize}
    \item Using Lemma 1.3, Lemma 1.4 directly implies that $\forall 0 \leq r < a_{2,|P|}, \forall S \subseteq P$ for which $|S^C| \leq \floor*{\frac{|P|}{|J|}r}$, if L1 uses the ordering strategy wrt $S$, then $|S \cap J| \geq |J| - r$. Note that $r + \floor*{\frac{|P|-|J|}{|J|}r} = \floor*{\frac{|P|}{|J|}r}$.

    \item Equivalently, $\forall a_{1,|P|} < r \leq |J|, \forall S \subseteq P$ for which $|S^C| \leq \floor*{\frac{|P|}{|J|}(|J| - r)}$, if L1 uses the ordering strategy wrt $S$, then $|S \cap J| \geq r$. This follows because $a_{1,|P|} < r \leq |J| \iff 0 \leq |J| - r < a_{2,|P|}$, because $|J| = a_{1,|P|} + a_{2,|P|}$.

    \item Thus, $\forall a_{1,|P|} < r \leq |J|, \forall S \subseteq P$ for which $|S| \geq |P| - \floor*{\frac{|P|}{|J|}(|J| - r)} = \ceil*{\frac{|P|}{|J|}r}$, if L1 uses the ordering strategy wrt $S$, then $|S \cap J| \geq r$. This is because $S$ and $S^C$ partition $P$, and so $|S| + |S^C| = |P|$. (Note that in general, $x - \floor*{y} = \ceil*{x-y}$).\\
\end{itemize}

Combining Corollaries 1.1 and 1.2, it follows that if eqs. (3) and (4) are true, then $\forall 0 \leq r \leq |J|, \forall S \subseteq P$ for which $|S| \geq \frac{|P|}{|J|}r$, if L1 uses the ordering strategy wrt $S$, then $|S \cap J| \geq r$. This is because $\forall x \in \mathbb{N}, y \in \mathbb{R}: x \geq y \iff x \geq \ceil{y}$. Noting that $|S| \leq |P|$, and therefore that there will never be $r > |J|$ for which $|S| \geq \frac{|P|}{|J|}r$, this means that $\forall S \subseteq P$, L1 can guarantee that $J$ is $(P,S)$-proportional by employing the ordering strategy wrt $S$. Thus, a $(|P|,|J|)$-HAP method that satisfies eqs. (3) and (4) is quasi-proportional wrt L1.

A similar proof establishes that if the inequalities opposite eqs. (3) and (4) are met, then the method will be quasi-proportional wrt L2.

\hfill$\blacksquare$\\

\textit{Intuition:} Theorem 1 sets upper and lower bounds, at every step of the procedure, on the ratio between (1) the number of affirmatives that a given litigant has exercised until that step and (2) the number of peremptories that the other litigant has exercised until that step. In particular, this ratio  of one litigant's affirmatives to the other's peremptories must always be close to $|J|:|P|-|J|$.

For instance, for $|P| = 100$ and $|J| = 10$, the following table shows, for each number of affirmatives that one litigant has exercised until a given step, how many peremptories the other litigant must have been able to exercise until that given step for the method to be quasi-proportional:

    \begin{center}
        \begin{tabular}{ |c||c|c|c|c|c|c|c|c|c|c|c| } 
         \hline
         Affs & 0 & 1 & 2 & 3 & 4 & 5 & 6 & 7 & 8 & 9 & 10 \\ 
         \hline
         Pers & 0-9 & 0-18 & 9-27 & 18-36 & 27-45 & 36-54 & 45-63 & 54-72 & 63-81 & 72-90 & 81-90 \\ 
         \hline
        \end{tabular}
    \end{center}

Intuitively, these bounds must be satisfied for a $(|P|,|J|)$-HAP method to be quasi-proportional so that, at no point, either litigant has so many peremptories that they ``overwhelm" the other litigant's affirmatives (as happens in a pure peremptory system) or has so many affirmatives they ``overwhelm" the other litigant's peremptories (as happens in a pure affirmative system).

\hfill

\textit{Theorem 2:} Any $(|P|,|J|)$-HAP method that does not satisfy eq. (2) is not quasi-proportional. That is, eq. (2) is not only a sufficient, but also a necessary, condition for a $(|P|,|J|)$-HAP method's quasi-proportionality.\\

\textit{Proof:} Let us consider a quasi-proportional $(|P|,|J|)$-HAP method. Consider any time $t$ and $S \subseteq P$ such that $|S| = a_{j,t} + b_{i,t}$. Because the method is quasi-proportional, $j$ must be able to guarantee that $|S \cap J| \geq \floor*{\frac{|J|}{|P|}|S|}$ (recall condition (1)). But Lemma 1.3 tells us that $i$ can guarantee that $|S \cap J| \leq a_{j,t}$. Thus, $a_{j,t} \geq \floor*{\frac{|J|}{|P|}|S|} = \floor*{\frac{|J|}{|P|}(a_{j,t} + b_{i,t})}$. Equivalently, $a_{j,t} > \frac{|J|}{|P|}(a_{j,t} + b_{i,t}) - 1$. Rearranging, one gets the right inequality of eq. (2).

In addition, $i$ must be able to guarantee that $|S^C \cap J| \geq \floor*{\frac{|J|}{|P|}(|P|-|S|)}$, or equivalently, that $|S \cap J| \leq |J| - \floor*{\frac{|J|}{|P|}(|P|-|S|)} = \ceil*{\frac{|J|}{|P|}|S|}$. But Lemma 1.1 tells us that $j$ can guarantee that $|S \cap J| \geq a_{j,t}$. Thus, $a_{j,t} \leq \ceil*{\frac{|J|}{|P|}|S|} = \ceil*{\frac{|J|}{|P|}(a_{j,t} + b_{i,t})}$ Equivalently, $a_{j,t} < \frac{|J|}{|P|}(a_{j,t} + b_{i,t}) + 1$. Rearranging, one gets the left inequality of eq. (2).

\hfill$\blacksquare$\\

\textit{Theorem 3:} A $(|P|,|J|)$-HAP method is quasi-proportional if and only if it is proportional.\\

\textit{Proof:} WLOG, I will prove that a $(|P|,|J|)$-HAP method is proportional wrt L1 \textit{if} eqs. (3) and (4) are met. Combining this with the opposite inequalities for L2 implies that eq. (2) implies proportionality. In turn, because eq. (2) is necessary and sufficient for quasi-proportionality, by Theorems 1 and 2, this yields the ``only if" portion of the above result. The ``if" portion is trivial.

Consider an arbitrary panel $P$ of size $|P|$, and a sequence of nested sets $\textbf{S} = \{S_i \subseteq P\}_{i=1}^I$. I will prove that if L1 adopts the following strategy (which I will call the ``ordering strategy wrt $\textbf{S}$"), then $J$ will be $(P, \textbf{S})$-proportional regardless of L2's strategy:

\begin{itemize}
    \item L1 orders the individuals in $P$ into a sequence $\{p_j\}_{j=1}^{|P|}$, such that $\forall S_i \in \textbf{S}$, for $1 \leq j \leq |S_i|$, $p_j \in S_i$. That is, L1 prioritizes ordering the members of $S_i$ first for all $i$. This is possible because of the nested nature of the $S_i$s. In particular, L1 can do this by ordering the members of $S_1$ first, then the members of $S_2 \setminus S_1$ ... (and then, generally, the members of $S_{i+1} \setminus S_i$ for incrementing $i$) ... and then finally the members of $P\setminus S_I$.
    \item Whenever L1 has an affirmative, use it on the \textit{first} available individual, i.e., the available individual with the smallest $i$.
    \item Whenever L1 has a peremptory, use it on the \textit{last} available individual, i.e., the available individual with the largest $i$.
\end{itemize}

Note that in using the ordering strategy wrt $\textbf{S}$, L1 is simultaneously using the ordering strategy wrt $S_i$ for all $S_i \in \textbf{S}$. Therefore, it directly follows from Corollaries 1.1 and 1.2 that if eqs. (3) and (4) hold, and L1 uses the ordering strategy wrt $\textbf{S}$, then $\forall S_i \in \textbf{S}$, the selected jury $J$ must be $(P,S_i)$-proportional. So by definition, if L1 uses the ordering strategy wrt $\textbf{S}$, then $J$ will be $(P,\textbf{S})$-proportional.

In other words, if eqs. (3) and (4) hold, then $\forall$ sequences of nested subsets $\textbf{S}$, L1 can guarantee that $J$ is $(P,\textbf{S})$-proportional by employing the ordering strategy wrt $\textbf{S}$. Thus, a $(|P|,|J|)$-HAP method that satisfies eqs. (3) and (4) is proportional wrt L1. A similar proof establishes that if the inequalities opposite eqs. (3) and (4) are met, then the method will be proportional wrt L2. Thus, a $(|P|,|J|)$-HAP method that meets eq. (2) is proportional.

\hfill$\blacksquare$\\

\textit{Theorem 4:} $\forall |P|, |J| \leq |P|, 0 \leq \alpha \leq |J|, \exists$ a proportional $(|P|,|J|)$-HAP method such that $a_{1,|P|} = \alpha$ and $a_{2,|P|} = |J| - \alpha$.\\

\textit{Proof:} For convenience, denote $\underline{\alpha} \equiv \frac{|P|-|J|}{|J|}\alpha - \frac{|P|}{|J|}$ and $\overline{\alpha} \equiv \frac{|P|-|J|}{|J|}\alpha + \frac{|P|}{|J|}$. Also, only consider $\alpha, \beta \in \mathbb{N}$.

Note that $\forall \alpha \in \mathbb{N},\exists \beta \in \mathbb{N} \cap (\underline{\alpha}, \overline{\alpha})$ because the range of the interval is at least 2.

Also note that for $\alpha_1 < \alpha_2$, it is the case that $\underline{\alpha_1} \leq \underline{\alpha_2}$ and $\overline{\alpha_1} \leq \overline{\alpha_2}$. Thus, 
\begin{align*}
    &\min\{\beta:\beta \in (\underline{\alpha_1}, \overline{\alpha_1})\} \leq \min\{\beta:\beta \in (\underline{\alpha_2}, \overline{\alpha_2})\} \text{, and }\\
    &\max\{\beta:\beta \in (\underline{\alpha_1}, \overline{\alpha_1})\} \leq \max\{\beta:\beta \in (\underline{\alpha_2}, \overline{\alpha_2})\} \text{.}
\end{align*}

Moreover, $\forall \alpha, \exists \beta$ s.t.
\begin{align*}
    \beta &\in (\underline{\alpha}, \overline{\alpha}) \cap (\underline{\alpha+1}, \overline{\alpha+1})\\
    &= (\underline{\alpha+1}, \overline{\alpha})\\
    &= (\frac{|P|-|J|}{|J|}(\alpha+1) - \frac{|P|}{|J|}, \frac{|P|-|J|}{|J|}\alpha + \frac{|P|}{|J|})\\
    &= (\frac{|P|-|J|}{|J|}\alpha - 1, \frac{|P|-|J|}{|J|}\alpha + \frac{|P|}{|J|}) \text{.}
\end{align*}
This is true because the range of this interval is $1+\frac{|P|}{|J|} \geq 2$ (because $\frac{|P|}{|J|} \geq 1$).\\

Finally, note that $\forall 0 \leq \alpha \leq |J|$, it is true that:
\begin{equation}
    \min\{\beta:\beta \in (\underline{\alpha}, \overline{\alpha})\} + \min\{\beta:\beta \in (\underline{|J| - \alpha}, \overline{|J| - \alpha})\} \leq |P| - |J|
\end{equation}
\begin{equation}
    \max\{\beta:\beta \in (\underline{\alpha}, \overline{\alpha})\} + \max\{\beta:\beta \in (\underline{|J| - \alpha}, \overline{|J| - \alpha})\} \geq |P| - |J|
\end{equation}
\indent
\textit{Proof:}
Recall that $\underline{\alpha} = \frac{|P|-|J|}{|J|}\alpha - \frac{|P|}{|J|}$ and $\underline{|J| - \alpha} = \frac{|P|-|J|}{|J|}(|J| - \alpha) - \frac{|P|}{|J|}$. Note that because $\beta \in \mathbb{N}$, $\min\{\beta:\beta \in (\underline{\alpha}, \overline{\alpha})\} \leq \underline{\alpha} + 1$ and $\min\{\beta:\beta \in (\underline{|J| - \alpha}, \overline{|J| - \alpha})\} \leq \underline{|J| - \alpha} + 1$. Here,

\begin{align*}
    \underline{\alpha} + 1 + \underline{|J| - \alpha} + 1 &= \frac{|P|-|J|}{|J|}\alpha - \frac{|P|}{|J|} + \frac{|P|-|J|}{|J|}(|J| - \alpha) - \frac{|P|}{|J|} + 2\\
    &= |P|-|J| - 2 \frac{|P|}{|J|} + 2\\
    &\leq |P| - |J|
\end{align*}
where the last inequality follows because $\frac{|P|}{|J|} \geq 1$. So eq. (5) must be true.

\hfill

\indent Also recall that $\overline{\alpha} = \frac{|P|-|J|}{|J|}\alpha + \frac{|P|}{|J|}$ and $\overline{|J| - \alpha} = \frac{|P|-|J|}{|J|}(|J| - \alpha) + \frac{|P|}{|J|}$. Note that because $\beta$ is always a natural number, $\max\{\beta:\beta \in (\underline{\alpha}, \overline{\alpha})\} \geq \overline{\alpha} - 1$ and $\max\{\beta:\beta \in (\underline{|J| - \alpha}, \overline{|J| - \alpha})\} \geq \overline{|J| - \alpha} - 1$. Here,

\begin{align*}
    \overline{\alpha} - 1 + \overline{|J| - \alpha} - 1 &= \frac{|P|-|J|}{|J|}\alpha + \frac{|P|}{|J|} + \frac{|P|-|J|}{|J|}(|J| - \alpha) + \frac{|P|}{|J|} - 2\\
    &= |P|-|J| + 2 \frac{|P|}{|J|} - 2\\
    &\geq |P| - |J|
\end{align*}
where the last inequality follows because $\frac{|P|}{|J|} \geq 1$. So eq. (6) must be true. 

\hfill

Combining all of the above facts, it is possible to construct a $(|P|,|J|)$-HAP method such that $a_{1,|P|} = \alpha$ and $a_{2,|P|} = |J| - \alpha$, and eq. (2) is met. This method would work by, $\forall t, i\neq j$, keeping $b_{i,t} \in (\underline{a_{j,t}}, \overline{a_{j,t}})$ (``the conditions"). It is possible to do this because:
\begin{enumerate}
    \item at $t=0$ (when $a_{1,0} = a_{2,0} = b_{1,0} = b_{2,0} = 0$), the conditions are met.
    
    \item Given some $a_{i}$, it is possible to incrementally increase $b_{j}$ to a point where $b_{j} \in (\underline{a_{i}},\overline{a_{i}}) \cap (\underline{a_{i}+1},\overline{a_{i}+1})$, at which point litigant $i$ can safely be given one more affirmative (to bring her up to $a_{i}+1$ affirmatives) without violating the conditions.
    
    \item Eq. (5) tells us that it is always possible to assign L1 and L2 \textit{enough} peremptories to keep the conditions met. That is, the number of peremptories that must be given so as not to violate the conditions does not exceed the total number of peremptories that will be given ($|P| - |J|$).
    
    \item Eq. (6) tells us that it is always possible to assign L1 and L2 \textit{few enough} peremptories to keep the conditions met. That is, the total number of peremptories that will be given ($|P| - |J|$) does not exceed the total number of peremptories that can be given without violating the conditions.
\end{enumerate}

\hfill$\blacksquare$

As one concrete example, if $|P| = 100$ and $|J| = 10$, if one wants to give L1 3 affirmatives and L2 7 affirmatives, it would be a proportional method to let L1 and L2 each exercise 9 peremptories, then let L1 and L2 each exercise 1 affirmative, and to repeat this process 2 more times for L1's affirmatives/L2's peremptories and 6 more times for L2's affirmatives/L1's peremptories.

\section{Droop Proportionality}

Until now, I have considered a notion of proportionality that can be broadly described as follows: a litigant should be able to guarantee that, for each whole ``block" of $\frac{|P|}{|J|}$ individuals they like in $P$, at least one of them is chosen to be in $J$. This fraction $\frac{|P|}{|J|}$ is similar to what is known in the electoral literature as the Hare quota (in an election, the total number of votes cast divided by the total number of seats to be filled).

However, this notion of proportionality might seem too weak. Consider a setting with arbitrary $P$ and $|J| = 1$. Then, for a litigant to be able to guarantee that the seated juror is in $S$, it must be that $S = P$ (else, $\floor*{\frac{|S|}{|P|}|J|} = 0$): that is, if the litigant dislikes even one potential juror, they cannot guarantee that they will like the seated juror. It might seem more natural to impose the stronger condition that a litigant should be able to guarantee that they like the seated juror so long as they like a \textit{majority} of $P$. 

Extending this intuition, one reaches the following general strengthening of the basic proportionality rule: a litigant should be able to guarantee that, for each whole ``block" of individuals they like in $P$ whose size exceeds $\frac{|P|}{|J|+1}$, at least one of them is chosen to be in $J$. This fraction $\frac{|P|}{|J|+1}$ is similar to what is known in the electoral literature as the Droop quota (in an election, the total number of votes cast divided by the total number of seats to be filled plus one). The basic idea behind the Droop quota is that it is the infimum of the set of numbers that fit into $P$ no more than $J$ times. Considering the above example where $|J| = 1$, the Droop quota is thus a simple majority, because $\frac{|P|}{2}$ is the infimum of all numbers that fit into $|P|$ no more than once.

Using this notion of proportionality based on the Droop quota, much of the previous analysis regarding necessary and sufficient conditions for proportionality, as well as existence, may be repeated.

Given a panel $P$ and a subset $S \subseteq P$, a jury $J$ will be deemed ``$(P,S)$-D-proportional" if 
\begin{equation}
    \forall q \in \mathbb{N} \text{ s.t. } |S| > \frac{|P|}{|J|+1}q \text{ (or, equivalently, } \frac{|J|+1}{|P|}|S| > q\text{), } |S \cap J| \geq q
\end{equation}

Note that condition (7) is equivalent to $|S \cap J| \geq \ceil*{\frac{|J|+1}{|P|}|S|}-1$, which in turn is equivalent to $|S \cap J| \geq \frac{|J|+1}{|P|}|S|-1$. This is because $\forall x \in \mathbb{N}, y \in \mathbb{R}: x \geq y \iff x \geq \ceil{y}$.

A jury selection method will be deemed ``quasi-$(|P|,|J|)$-D-proportional wrt litigant $i$" if $\forall P$ of cardinality $|P|$, and $S \subseteq P$, $i$ can guarantee that the seated jury $J$ (of size $|J|$) will be $(P,S)$-D-proportional. A jury selection method will be deemed ``quasi-$(|P|,|J|)$-D-proportional" if it is quasi-$(|P|,|J|)$-D-proportional wrt both litigants.

Given a panel $P$ and a sequence of nested subsets $\textbf{S} = \{S_i \subseteq P\}_{i=1}^I$ such that $S_i \subset S_{i+1}$ for all $i < I$, the selected jury $J$ will be deemed ``$(P,\textbf{S})$-D-proportional" if $\forall S_i \in \textbf{S}$, $J$ is $(P,S_i)$-D-proportional.

A jury selection method will be deemed ``$(|P|,|J|)$-D-proportional wrt litigant $i$" if $\forall P$ of cardinality $|P|$, and sequence of nested subsets $\textbf{S}$, $i$ can guarantee that the seated jury $J$ (of size $|J|$) is $(P,\textbf{S})$-D-proportional.

A jury selection method will be deemed ``$(|P|,|J|)$-D-proportional" if it is $(|P|,|J|)$-D-proportional wrt both litigants.

Note that in both senses of the terms, $(|P|,|J|)$-D-proportionality implies quasi-$(|P|,|J|)$-D-proportionality.

\hfill

\textit{Theorem 5:} Any $(|P|,|J|)$-HAP method that has the following properties is quasi-D-proportional.

\begin{align}
    \forall i,j \in \{1,2\} (i \neq j), \forall t: \frac{|P|-|J|-1}{|J|+1}a_{j,t} \leq b_{i,t} \leq \frac{|P|-|J|-1}{|J|+1}a_{j,t} + \frac{|P|}{|J|+1}
\end{align}

\textit{Proof:} WLOG, I will prove that a $(|P|,|J|)$-HAP method is quasi-D-proportional wrt L1 \textit{if}

\begin{align}
    \forall t, b_{1,t} &\geq \frac{|P|-|J|-1}{|J|+1}a_{2,t} \\    
    \forall t, b_{2,t} &\leq \frac{|P|-|J|-1}{|J|+1}a_{1,t} + \frac{|P|}{|J|+1}
\end{align}

Combining this with the opposite inequalities for L2 yields the above result.\\

Consider an arbitrary panel $P$ of size $|P|$ and subset $S \subseteq P$, where a jury $J\subseteq P$ of size $|J|$ must be chosen. I will prove that if L1 adopts the ordering strategy wrt $S$, then $J$ will be $(P,S)$-D-proportional regardless of L2's strategy.\\

\textit{Lemma 1.1:} If $\exists t$ and $x,y \in \mathbb{N}$ s.t. $a_{1,t} = x$ and $b_{2,t} \leq y$, then $\forall S\subseteq P$ for which $|S| \geq x+y$, if L1 uses the ordering strategy wrt $S$, then $|S \cap J| \geq x$. In other words, if there is any time $t$ at which litigant 1 has $x$ affirmatives while litigant 2 has at most $y$ peremptories, then for any group $S$ which contains at least $x+y$ individuals, L1 can guarantee that at least $x$ of them end up on the final jury by employing the ordering strategy wrt $S$.\\

\textit{Lemma 5.2:} Assuming eq. (10) is true, then $\forall r \leq a_{1,|P|}, \exists t$ s.t. $a_{1,t} = r$ and $b_{2,t} \leq \floor*{\frac{|P|-|J|-1}{|J|+1}r}+1$. That is, for any number no greater than the total number of affirmatives that L1 gets, there will be a time at which L1 has that many affirmatives \textit{and} L2 has no greater than a threshold number of peremptories.\\

\textit{Proof:} First note that this is trivially true for $r=0$, because the inequality would be satisfied at $t=0$ (because $b_{2,0} = 0$).

For $r>0$, I prove this lemma by contrapositive, by showing that if it doesn't hold, then eq. (10) can't be true.

Suppose $\exists \text{ } 1 \leq r \leq a_{1,|P|}$ s.t. $\forall t$ for which $a_{1,t} = r$, $b_{2,t} > \frac{|P|-|J|-1}{|J|+1}a_{1,t}+1$. Because this must hold for all $t$ for which $a_{1,t} = r$, it must hold in particular for the $t'$ at which $a_{1,t'-1} = a_{1,t'}-1 = r-1$ (that is, for the first $t'$ at which L1 had $r$ affirmatives). Define $t''\equiv t'-1$ (the last $t$ at which L1 had $r-1$ affirmatives). Note that because only one action can happen at each time, it must be that $b_{2,t''} = b_{2,t'}$ (that is, L2 still had the same number of peremptories at $t''$ as she did at $t'$).

In this case, 
\begin{align*}
    \frac{|P|-|J|-1}{|J|+1}a_{1,t''}+\frac{|P|}{|J|+1} &= \frac{|P|-|J|-1}{|J|+1}(a_{1,t'}-1)+\frac{|P|}{|J|+1}\\
    &= \frac{|P|-|J|-1}{|J|+1}a_{1,t'}+1\\
    &< b_{2,t'} = b_{2,t''}
\end{align*}

Note that this inequality \textit{violates} eq. (10). Thus, by contrapositive, if eq. (10) is true, then $\forall r \leq a_{1,|P|}, \exists t$ s.t. $a_{1,t} = r$ and $b_{2,t} \leq \frac{|P|-|J|-1}{|J|+1}r+1$. And $\forall x \in \mathbb{N}, y \in \mathbb{R}: x \leq y \iff x \leq \floor{y}$.\\

\textit{Corollary 5.1:} If eq. (10) is true, then Lemmas 1.1 and 5.2 together imply that $\forall 0 \leq r \leq a_{1,|P|}, \forall S \subseteq P$ for which $|S| \geq \floor*{\frac{|P|}{|J|+1}r}+1$, L1 can guarantee that $|S \cap J| \geq r$ by using the ordering strategy wrt $S$. Note that $r + \floor*{\frac{|P|-|J|-1}{|J|+1}r} + 1 = \floor*{\frac{|P|}{|J|+1}r}+1$.

\hfill

\textit{Lemma 1.3:} If $\exists t$ and $x,y \in \mathbb{N}$ s.t. $a_{2,t} = x$ and $b_{1,t} \geq y$, then $\forall S\subseteq P$ for which $|S^C| \leq x+y$, if L1 uses the ordering strategy wrt $S$, then $|S \cap J| \geq |J| - x$. In other words, if there is any time $t$ at which L2 has $x$ affirmatives while L1 has at least $y$ peremptories, then for any group $S^C$ which contains at most $x+y$ individuals, L1 guarantees that at most $x$ of them end up on the final jury by employing the ordering strategy wrt $S$.\\

\textit{Lemma 5.4:} Assuming eq. (9) is true, then $\forall r < a_{2,|P|}, \exists t$ s.t. $a_{2,t} = r$ and $b_{1,t} \geq \ceil*{\frac{|P|-|J|-1}{|J|+1}(r+1)}$. That is, for any number less than the total number of affirmatives that L2 gets, there will be a time at which L2 has that many affirmatives \textit{and} L1 has no fewer than a threshold number of peremptories.\\

\textit{Proof:} For $0\leq r<a_{2,|P|}$, I prove this lemma by contrapositive, by showing that if it doesn't hold, then eq. (9) can't be true.

Suppose $\exists \text{ } 0 \leq r < a_{2,|P|}$ s.t. $\forall t$ for which $a_{2,t} = r$, $b_{1,t} < \frac{|P|-|J|-1}{|J|+1}(a_{2,t}+1)$. Because this must hold for all $t$ for which $a_{2,t} = r$, it must hold in particular for the $t'$ at which $a_{2,t'+1} = a_{2,t'}+1 = r+1$ (that is, for the last $t'$ at which L2 had $r$ affirmatives). Define $t''\equiv t'+1$ (the first $t$ at which L2 had $(r+1)$ affirmatives). Note that because only one action can happen at each time, it must be that $b_{1,t''} = b_{1,t'}$ (that is, L1 still had the same number of peremptories at $t'$ as he did at $t''$).

In this case, 
\begin{align*}
    \frac{|P|-|J|-1}{|J|+1}a_{2,t''} &= \frac{|P|-|J|-1}{|J|+1}(a_{2,t'}+1)\\
    &> b_{1,t'} = b_{1,t''}
\end{align*}

Note that this inequality \textit{violates} eq. (9). Thus, by contrapositive, if eq. (9) is true, then $\forall r < a_{2,|P|}, \exists t$ s.t. $a_{2,t} = r$ and $b_{1,t} \geq \frac{|P|-|J|-1}{|J|+1}(r+1)$. And $\forall x \in \mathbb{N}, y \in \mathbb{R}: x \geq y \iff x \geq \ceil{y}$.\\

\textit{Corollary 5.2:} If eq. (9) is true, then $\forall a_{1,|P|} < r \leq |J|, \forall S \subseteq P$ for which $|S| \geq \floor*{\frac{|P|}{|J|+1}r}+1$, if L1 uses the ordering strategy wrt $S$, then $|S \cap J| \geq r$.

\textit{Proof:}
\begin{itemize}
    \item Using Lemma 1.3, Lemma 5.4 directly implies that $\forall 0 \leq r < a_{2,|P|}, \forall S \subseteq P$ for which $|S^C| \leq \ceil*{\frac{|P|}{|J|+1}(r+1)}-1$, if L1 uses the ordering strategy wrt $S$, then $|S \cap J| \geq |J| - r$. Note that $r + \ceil*{\frac{|P|-|J|-1}{|J|+1}(r+1)} = \ceil*{\frac{|P|}{|J|+1}(r+1)}-1$.

    \item Equivalently, $\forall a_{1,|P|} < r \leq |J|, \forall S \subseteq P$ for which $|S^C| \leq \ceil*{\frac{|P|}{|J|+1}(|J|-r+1)} - 1$, if L1 uses the ordering strategy wrt $S$, then $|S \cap J| \geq r$. This follows because $a_{1,|P|} < r \leq |J| \iff 0 \leq |J| - r < a_{2,|P|}$, because $|J| = a_{1,|P|} + a_{2,|P|}$.

    \item Thus, $\forall a_{1,|P|} < r \leq |J|, \forall S \subseteq P$ for which $|S| \geq |P| - \ceil*{\frac{|P|}{|J|+1}(|J|-r+1)} + 1 = \floor*{\frac{|P|}{|J|+1}r}+1$, if L1 uses the ordering strategy wrt $S$, then $|S \cap J| \geq r$. This is because $S$ and $S^C$ partition $P$, and so $|S| + |S^C| = |P|$.\\
\end{itemize}

Combining Corollaries 5.1 and 5.2, it follows that if eqs. (9) and (10) are true, then $\forall 0 \leq r \leq |J|, \forall S \subseteq P$ for which $|S| > \frac{|P|}{|J|+1}r$, if L1 uses the ordering strategy wrt $S$, then $|S \cap J| \geq r$. This is because $\forall x \in \mathbb{N}, y \in \mathbb{R}: x > y \iff x \geq \floor{y}+1$. Noting that $|S| \leq |P|$, and therefore that there will never be $r > |J|$ for which $|S| > \frac{|P|}{|J|+1}r$, this means that $\forall S \subseteq P$, L1 can guarantee that $J$ is $(P,S)$-D-proportional by employing the ordering strategy wrt $S$. Thus, a $(|P|,|J|)$-HAP method that satisfies eqs. (9) and (10) is quasi-D-proportional wrt L1.

A similar proof establishes that if the inequalities opposite eqs. (9) and (10) are met, then the method will be quasi-D-proportional wrt L2.

\hfill$\blacksquare$\\

\textit{Theorem 6:} Any $(|P|,|J|)$-HAP method that does not satisfy eq. (8) is not quasi-D-proportional. That is, eq. (8) is not only a sufficient, but also a necessary, condition for a $(|P|,|J|)$-HAP method's quasi-D-proportionality.\\

\textit{Proof:} Let us consider a quasi-D-proportional $(|P|,|J|)$-HAP method. Consider any time $t$ and $S \subseteq P$ such that $|S| = a_{j,t} + b_{i,t}$. Because the method is quasi-D-proportional, $j$ must be able to guarantee that $|S \cap J| \geq \frac{|J|+1}{|P|}|S|-1$ (recall condition (7)). But Lemma 1.3 tells us that $i$ can guarantee that $|S \cap J| \leq a_{j,t}$. Thus, $a_{j,t} \geq \frac{|J|+1}{|P|}|S|-1 = \frac{|J|+1}{|P|}(a_{j,t} + b_{i,t})-1$. Rearranging, one gets the right inequality of eq. (8).

In addition, $i$ must be able to guarantee that $|S^C \cap J| \geq \frac{|J|+1}{|P|}(|P|-|S|) - 1$, or equivalently, that $|S \cap J| \leq |J| - \frac{|J|+1}{|P|}(|P|-|S|) + 1 = \frac{|J|+1}{|P|}|S|$. But Lemma 1.1 tells us that $j$ can guarantee that $|S \cap J| \geq a_{j,t}$. Thus, $a_{j,t} \leq \frac{|J|+1}{|P|}|S| = \frac{|J|+1}{|P|}(a_{j,t} + b_{i,t})$. Rearranging, one gets the left inequality of eq. (8).

\hfill$\blacksquare$\\

\textit{Theorem 7:} A $(|P|,|J|)$-HAP method is quasi-D-proportional if and only if it is D-proportional.\\

\textit{Proof:} WLOG, I will prove that a $(|P|,|J|)$-HAP method is D-proportional wrt L1 \textit{if} eqs. (9) and (10) are met. Combining this with the opposite inequalities for L2 implies that eq. (8) implies D-proportionality. In turn, because eq. (8) is necessary and sufficient for quasi-D-proportionality, by Theorems 5 and 6, this yields the ``only if" portion of the above result. The ``if" portion is trivial.

Consider an arbitrary panel $P$ from which a jury $J\subseteq P$ must be chosen, and a sequence of nested sets $\textbf{S} = \{S_i \subseteq P\}_{i=1}^I$. I will prove that if L1 adopts the following strategy (which I will call the ``ordering strategy wrt $\textbf{S}$"), then $J$ will be $(P, \textbf{S})$-D-proportional regardless of L2's strategy:

\begin{itemize}
    \item L1 orders the individuals in $P$ into a sequence $\{p_j\}_{j=1}^{|P|}$, such that $\forall S_i \in \textbf{S}$, for $1 \leq j \leq |S_i|$, $p_j \in S_i$. That is, L1 prioritizes ordering the members of $S_i$ first for all $i$. This is possible because of the nested nature of the $S_i$s. In particular, L1 can do this by ordering the members of $S_1$ first, then the members of $S_2 \setminus S_1$ ... (and then, generally, the members of $S_{i+1} \setminus S_i$ for incrementing $i$) ... and then finally the members of $P\setminus S_I$.
    \item Whenever L1 has an affirmative, use it on the \textit{first} available individual, i.e., the available individual with the smallest $i$.
    \item Whenever L1 has a peremptory, use it on the \textit{last} available individual, i.e., the available individual with the largest $i$.
\end{itemize}

Note that in using the ordering strategy wrt $\textbf{S}$, L1 is simultaneously using the ordering strategy wrt $S_i$ for all $S_i \in \textbf{S}$. Therefore, it directly follows from Corollaries 5.1 and 5.2 that if eqs. (9) and (10) hold, and L1 uses the ordering strategy wrt $\textbf{S}$, then $\forall S_i \in \textbf{S}$, the selected jury $J$ must be $(P,S_i)$-D-proportional. So by definition, if L1 uses the ordering strategy wrt $\textbf{S}$, then $J$ will be $(P,\textbf{S})$-D-proportional.

In other words, if eqs. (9) and (10) hold, $\forall$ sequences of nested subsets $\textbf{S}$, L1 can guarantee that $J$ is $(P,\textbf{S})$-D-proportional by employing the ordering strategy wrt $\textbf{S}$. Thus, a $(|P|,|J|)$-HAP method that satisfies eqs. (9) and (10) is D-proportional wrt L1. A similar proof establishes that if the inequalities opposite eqs. (9) and (10) are met, then the method will be D-proportional wrt L2. Thus, a $(|P|,|J|)$-HAP method that meets eq. (8) is D-proportional.

\hfill$\blacksquare$\\

\textit{Theorem 8:} $\forall |P|, |J| \leq |P|, 0 \leq \alpha \leq |J|, \exists$ a D-proportional $(|P|,|J|)$-HAP method such that $a_{1,|P|} = \alpha$ and $a_{2,|P|} = |J| - \alpha$.\\

\textit{Proof:}
The result is trivial when $|P| = |J|$, as then everyone on the panel is chosen: there are no peremptories, and the affirmatives may be distributed between the two litigants in any way. I next consider when $|P| > |J|$.

For convenience, denote $\underline{\alpha} \equiv  \frac{|P|-|J|-1}{|J|+1}\alpha$ and $\overline{\alpha} \equiv \frac{|P|-|J|-1}{|J|+1}\alpha + \frac{|P|}{|J|+1}$. Also, only consider $\alpha, \beta \in \mathbb{N}$.

Note that $\forall \alpha \in \mathbb{N},\exists \beta \in \mathbb{N} \cap [\underline{\alpha}, \overline{\alpha}]$ because the range of the interval is at least 1 and the interval is closed.

Also note that for $\alpha_1 < \alpha_2$, it is the case that $\underline{\alpha_1} \leq \underline{\alpha_2}$ and $\overline{\alpha_1} \leq \overline{\alpha_2}$, from the definition of $\underline{\alpha}$ and $\overline{\alpha}$. Thus, 
\begin{align*}
    &\min\{\beta:\beta \in [\underline{\alpha_1}, \overline{\alpha_1}]\} \leq \min\{\beta:\beta \in [\underline{\alpha_2}, \overline{\alpha_2}]\} \text{, and }\\
    &\max\{\beta:\beta \in [\underline{\alpha_1}, \overline{\alpha_1}]\} \leq \max\{\beta:\beta \in [\underline{\alpha_2}, \overline{\alpha_2}]\} \text{.}
\end{align*}

Moreover, $\forall \alpha \in \mathbb{N}, \exists \beta \in \mathbb{N}$ s.t. 
\begin{align*}
    \beta &\in [\underline{\alpha}, \overline{\alpha}] \cap [\underline{\alpha+1}, \overline{\alpha+1}]\\
    &= [\underline{\alpha+1}, \overline{\alpha}]\\
    &= [\frac{|P|-|J|-1}{|J|+1}(\alpha+1),  \frac{|P|-|J|-1}{|J|+1}\alpha + \frac{|P|}{|J|+1}]\\
    &= [\frac{|P|-|J|-1}{|J|+1}\alpha + \frac{|P|}{|J|+1}-1, \frac{|P|-|J|-1}{|J|+1}\alpha + \frac{|P|}{|J|+1}] \text{.}
\end{align*}
This is true because the range of this interval is 1 and the interval is closed.\\

Finally, note that $\forall 0 \leq \alpha \leq |J|$, it is true that:
\begin{equation}
    \min\{\beta:\beta \in [\underline{\alpha}, \overline{\alpha}]\} + \min\{\beta:\beta \in [\underline{|J| - \alpha}, \overline{|J| - \alpha}]\} = \ceil{\underline{\alpha}} + \ceil{\underline{|J| - \alpha}} \leq |P| - |J|
\end{equation}
\begin{equation}
    \max\{\beta:\beta \in [\underline{\alpha}, \overline{\alpha}]\} + \max\{\beta:\beta \in [\underline{|J| - \alpha}, \overline{|J| - \alpha}]\} = \floor{\underline{\alpha}} + \floor{\underline{|J| - \alpha}} \geq |P| - |J|
\end{equation}
\indent
\textit{Proof:}
Recall that $\underline{\alpha} = \frac{|P|-|J|-1}{|J|+1}\alpha$ and $\underline{|J| - \alpha} =  \frac{|P|-|J|-1}{|J|+1}(|J|-\alpha)$. Here,

\begin{align*}
    \underline{\alpha} + \underline{|J| - \alpha} &=  \frac{|P|-|J|-1}{|J|+1}|J| = \frac{|J|}{|J|+1}(|P|-|J|-1)\\
    &\leq |P| - |J| - 1
\end{align*}
In addition, note that if $x+y \leq z$, where $z \in \mathbb{N}$, then $\ceil{x} + \ceil{y} \leq z + 1$. (Proof: $\ceil{x}-1 < x, \ceil{y}-1 < y \implies \ceil{x} + \ceil{y} < x + y + 2 \leq z + 2 \implies \ceil{x} + \ceil{y} \leq z + 1$ (because $\ceil{x}$, $\ceil{y}$, and $z$ are integers).) So eq. (11) must be true.

\hfill

\indent Also recall that $\overline{\alpha} =  \frac{|P|-|J|-1}{|J|+1}\alpha + \frac{|P|}{|J|+1}$ and $\overline{|J| - \alpha} = \frac{|P|-|J|-1}{|J|+1}(|J|-\alpha) + \frac{|P|}{|J|+1}$. Here,

\begin{align*}
    \overline{\alpha} + \overline{|J| - \alpha} &= \frac{|P|-|J|-1}{|J|+1}|J| + 2\frac{|P|}{|J|+1} = \frac{|J|+2}{|J|+1}|P| - |J|\\
    & = \frac{|P|}{|J|+1} + |P| - |J| \geq |P| - |J| + 1
\end{align*}
where the last inequality follows because $|P| \geq |J| + 1$. In addition, note that if $x + y \geq z$, where $z \in \mathbb{N}$, then $\floor{x} + \floor{y} \geq z - 1$. (Proof: $\floor{x}+1 > x, \floor{y}+1 > y \implies \floor{x} + \floor{y} > x + y - 2 \geq z - 2 \implies \floor{x} + \floor{y} \geq z - 1$ (because $\floor{x}$, $\floor{y}$, and $z$ are integers).) So eq. (12) must be true. 

\hfill

Combining all of the above facts, it must be true that there exists a $(|P|,|J|)$-HAP method such that $a_{1,|P|} = \alpha$ and $a_{2,|P|} = |J| - \alpha$, and eq. (8) is met. This method would work by, at $\forall t, i\neq j$, keeping $b_{i,t} \in [\underline{a_{j,t}}, \overline{a_{j,t}}]$ (``the conditions"). It is possible to do this because:
\begin{enumerate}
    \item At $t=0$ (when $a_{1,0} = a_{2,0} = b_{1,0} = b_{2,0} = 0$), the conditions are met.
    
    \item Given some $a_{i}$, it is possible to incrementally increase $b_{j}$ to a point where $b_{j} \in [\underline{a_{i}},\overline{a_{i}}] \cap [\underline{a_{i}+1},\overline{a_{i}+1}]$, at which point litigant $i$ can safely be given one more affirmative (to bring her up to $a_{i}+1$ affirmatives) without violating the conditions.
    
    \item Eq. (11) tells us that it is always possible to assign L1 and L2 \textit{enough} peremptories to keep the conditions met. That is, the number of peremptories that must be given so as not to violate the conditions does not exceed the total number of peremptories that will be given ($|P| - |J|$).
    
    \item Eq. (12) tells us that it is always possible to assign L1 and L2 \textit{few enough} peremptories to keep the conditions met. That is, the total number of peremptories that will be given ($|P| - |J|$) does not exceed the total number of peremptories that can be given without violating the conditions.
\end{enumerate}

\hfill$\blacksquare$

\section{Proportionality and Representativeness}

One might think of a ``representative" $J$ as one whose distribution (wrt whatever metric a litigant cares about) closely approximates, or is ``better" than, that of $P$. The following result shows that this conception is intertwined with the previously formulated conception of ``proportionality."\\

\textit{Theorem 9:} Consider the CDF $F_P$ of an arbitrary $P$ wrt a metric that is increasing in litigant $i$'s preference. If a $(|P|,|J|)$-proportional jury selection method is employed, then $i$ can guarantee that the CDF $F_J$ of $J$ (of size $|J|$) wrt that metric nowhere exceeds $F_P$ by more than $\frac 1 {|J|}$. That is, $i$ can guarantee that $\forall x, F_J(x) - F_P(x) < \frac 1 {|J|}$.\\

\textit{Proof:} Consider arbitrary $P$ of size $|P|$. Recall that by the definition of a $(|P|,|J|)$-proportional method, $i$ should be able to guarantee that $\forall$ sequences of nested subsets $\textbf{S} = \{S_k \subseteq P\}_{k=1}^K$, and $\forall S_k \in \textbf{S}$, $|S_k \cap J| \geq \floor*{\frac{|J|}{|P|}|S_k|}$. Let $i$ choose $\textbf{S}$ to reflect her preferences, so that $S_1$ comprises her most preferred individuals, $S_2$ comprises her most and second-most preferred individuals, and so on. Consider the jury $J$ that results.

Consider any value $x$ of the relevant metric. Consider the set of members of $P$ for whom the value of the relevant metric is $>x$ (note: this is the subset of $P$ that $i$ sufficiently likes). Call this set $S'$. Note that $S' \in \textbf{S}$ because $\textbf{S}$ reflects $i$'s preferences. Thus, $i$ can guarantee that $|S' \cap J| \geq \floor*{\frac{|J|}{|P|}|S'|}$.

Note that $|S' \cap J| \geq \floor*{\frac{|J|}{|P|}|S'|} \iff |S' \cap J| > \frac{|J|}{|P|} |S'| - 1$. Dividing by $|J|$, $\frac{|S' \cap J|}{|J|} > \frac{|S'|}{|P|} - \frac{1}{|J|}$. This, in turn, means that $\frac{|S'^C \cap J|}{|J|} < \frac{|S'^C|}{|P|} + \frac{1}{|J|}$ (because $|S' \cap J| + |S'^C \cap J| = |J|$ and $|S'| + |S'^C| = |P|$).

Finally, noting that $F_J(x) = \frac{|S'^C \cap J|}{|J|}$ and $F_P(x) = \frac{|S'^C|}{|P|}$, I get that for the chosen $x$, $F_J(x) - F_P(x) < \frac 1 {|J|}$. Because $x$ was chosen arbitrarily, this is true for all $x$.

\hfill$\blacksquare$

\textit{Corollary:} As $|J| \to \infty$, under a $(|P|,|J|)$-proportional jury selection method, litigant $i$ can guarantee that wrt their preferred metric, $F_J$ first order stochastically dominates $F_P$.

\hfill

\textit{Theorem 10:} Consider the CDF $F_P$ of an arbitrary $P$ wrt a metric that is increasing in litigant $i$'s preference. If a $(|P|,|J|)$-D-proportional jury selection method is employed, then $i$ can guarantee that the CDF $F_J$ of $J$ (of size $|J|$) wrt that metric exceeds $F_P$ by no more than $\frac 1 {|J|+1}$ anywhere. That is, $i$ can guarantee that $\forall x, F_J(x) - F_P(x) \leq \frac 1 {|J|+1}$.\\

\textit{Proof:} Consider arbitrary $P$ of size $|P|$. Recall that by the definition of a $(|P|,|J|)$-D-proportional method, $i$ should be able to guarantee that $\forall$ sequences of nested subsets $\textbf{S} = \{S_k \subseteq P\}_{k=1}^K$, and $\forall S_k \in \textbf{S}$, $|S_k \cap J| \geq \frac{|J|+1}{|P|}|S_k|-1$. Let $i$ choose $\textbf{S}$ to reflect her preferences, so that $S_1$ comprises her most preferred individuals, $S_2$ comprises her most and second-most preferred individuals, and so on. Consider the jury $J$ that results.

Consider any value $x$ of the relevant metric. Consider the set of members of $P$ for whom the value of the relevant metric is $>x$ (note: this is the subset of $P$ that $i$ sufficiently likes). Call this set $S'$. Note that $S' \in \textbf{S}$ because $\textbf{S}$ reflects $i$'s preferences. Thus, $i$ can guarantee that $|S' \cap J| \geq \frac{|J|+1}{|P|}|S'|-1$.

Dividing $|S' \cap J| \geq \frac{|J|+1}{|P|}|S'|-1$ by $|J|$, we get $\frac{|S' \cap J|}{|J|} \geq \frac{|S'|}{|P|} + \frac{|S'|}{|P|}\frac{1}{|J|} - \frac{1}{|J|}$. This, in turn, means that $\frac{|S'^C \cap J|}{|J|} \leq \frac{|S'^C|}{|P|} + \frac{1}{|J|}(1-\frac{|S'|}{|P|}) = \frac{|S'^C|}{|P|} + \frac{1}{|J|}\frac{|S'^C|}{|P|}$. Further noting that $F_J(x) = \frac{|S'^C \cap J|}{|J|}$ and $F_P(x) = \frac{|S'^C|}{|P|}$, we get that, for this value of $x$, $F_J(x) \leq \frac{|J|+1}{|J|}F_P(x)$. Because $x$ was chosen arbitrarily, this is true for all $x$.

Because $F_J(x)$ cannot exceed 1, we have that $F_J(x) \leq \min\{1,\frac{|J|+1}{|J|}F_P(x)\}$. Thus, $F_J(x) - F_P(x) \leq \min\{1 - F_P(x),\frac{1}{|J|}F_P(x)\}$. This expression reaches its maximum value of $\frac 1 {|J|+1}$ when $F_P(x) = \frac{|J|}{|J|+1}$. Thus, $\forall x, F_J(x) - F_P(x) \leq \frac 1 {|J|+1}$.

\hfill$\blacksquare$

\section{Eliciting preferences from litigants}

One way to implement an HAP method is to go through the process manually, by asking the appropriate litigant at each step which available member of $P$ they would like to exercise their action on. Under this implementation, a litigant would know what has happened in previous steps when they make their decision at a given step.

An alternative is to have each litigant submit their preference ordering (permitting indifference) to the judge, and for the judge to employ the following procedure:

\begin{itemize}
    \item The judge, for each litigant $i$, orders the  individuals in $P$ into a sequence $\{p_k\}_{k=1}^{|P|}$, such that the individuals whom $i$ more prefers have lower indices. They do this in such a way that for individuals among whom $i$ is indifferent, individuals whom $j$ prefers have lower indices.

    \item In a step where a litigant is assigned an affirmative, use it on the available individual with the smallest index $k$ according to the constructed ordering for that litigant.
    
    \item In a step where a litigant is assigned a peremptory, use it on the available individual with the largest index $k$ according to the constructed ordering for that litigant.
\end{itemize}

That is, the litigants would submit their preferences up front and the judge would mechanically apply a sort of ordering strategy wrt each litigant's preferences. I will call this procedure ``ordering wrt elicited preferences."

\hfill

\textit{Proposition 11:} Ordering wrt elicited preferences preserves the (D-)proportionality of a $(|P|,|J|)$-HAP method.

\textit{Proof:} $\forall P$ of size $|P|$, litigant $i$ is able to guarantee that the seated jury $J$ is (D-)proportional wrt any sequence of nested subsets $\textbf{S} = \{S_l \subseteq P\}_{l=1}^L$ by submitting a preference ordering that reflects that they most prefer, and are indifferent among, elements of $S_1$; that they second-most prefer, and are indifferent among, elements of $S_2 \setminus S_1$; and so on. This is because the judge, by ordering wrt elicited preferences, then mechanically employs the ordering strategy wrt $\textbf{S}$ (note that the specific ordering of individuals who are elements of the same ``layers" of $\textbf{S}$ (i.e. individuals among whom $i$ is indifferent) does not matter, so the judge is permitted to consider $j$'s preferences in ordering them).

\hfill

\textit{Proposition 12:} Ordering wrt elicited preferences is juror-wise Pareto efficient wrt the elicited preferences. That is, if both L1 and L2 weakly prefer Alice to Bob, with one preference strict, then ordering wrt elicited preferences will never select Bob but not Alice to sit on $J$.

\textit{Proof:} Suppose litigants $i$ and $j$ both report that they weakly prefer Alice to Bob, with at least one preference being strict. Then, by construction, the judge's ordering of $P$ for both litigants will give Alice a lower index than Bob. Thus, affirmatives of both litigants would be exercised on Alice before Bob and peremptories of both litigants would be exercised on Bob before Alice. This means it would be impossible for Bob, but not Alice, to be on $J$.

\hfill

\textit{Proposition 13:} Ordering wrt elicited preferences according to a (D-)proportional $(|P|,|J|)$-HAP method is not necessarily a strategy-proof jury selection method. That is, it would not be a dominant strategy for litigants to honestly report their preferences.

\textit{Proof:} Consider $P = \{A,B,C\}$ and $|J| = 1$. Suppose L1's preferences are $A \succ B \succ C$ and L2's are $B \succ A \succ C$.

Consider a (D-proportional) $(3,1)$-HAP method wherein:
\begin{itemize}
    \item at $t=1$: L1 exercises a peremptory
    
    \item at $t=2$: L2 exercises a peremptory
    
    \item at $t=3$: L1 exercises an affirmative
\end{itemize}

If L1 knows that L2 is going to report their preferences honestly, then L1 would prefer to report their own preferences as $A \succ C \succ B$. If they were to report honestly: $C$ would be eliminated, then $A$ would be eliminated, then $B$ would be chosen. But if they were to misreport as specified: $B$ would be eliminated, then $C$ would be eliminated, then $A$ would be chosen.

\section{More than two litigants}
\textit{Proposition 14:} There does not exist a jury selection method that is quasi-proportional wrt each of $N \geq 6$ litigants.

\textit{Proof:} Consider $P = \{A,B,C,D\}$ and $|J|=2$. L1 likes $\{A,B\}$; L2 likes $\{A,C\}$; L3 likes $\{A,D\}$; L4 likes $\{B,C\}$; L5 likes $\{B,D\}$; L6 likes $\{C,D\}$. Because each litigant likes two members of $P$ out of four, they should be able to guarantee that they like at least one of the two members of $J$. But no matter which two members of $P$ are chosen, there will be at least one litigant who likes neither member. So not all litigants are able to guarantee that $J$ is proportional wrt their $S$.

\hfill

\textit{Proposition 15:} There does not exist a jury selection method that is quasi-D-proportional wrt each of $N \geq 3$ litigants.

\textit{Proof:} Consider $P = \{A,B,C,D,E\}$ and $|J|=1$. L1 likes $\{A,B,C\}$; L2 likes $\{C,D,E\}$; L3 likes $\{A,B,D\}$. Because each litigant likes three members of $P$ out of five (i.e. $>\frac12$), they should be able to guarantee that they like the chosen member of $J$. But no matter which member of $P$ is chosen, there will be at least one litigant who does not like them. So not all litigants are able to guarantee that $J$ is D-proportional wrt their $S$.

\hfill

\textit{Proposition 16:} There does not always exist an HAP method that is quasi-proportional wrt each of $N \geq 3$ litigants.

\textit{Proof:} Let $|P| = 6$ and $|J| = 3$. Consider some number of litigants $N \geq 3$. Note that for an HAP method to be quasi-proportional wrt a litigant, it must be that:

\begin{align*}
    \forall t, b_{i,t} > \frac{|P|-|J|}{|J|}\sum_{j \neq i} a_{j,t} - \frac{|P|}{|J|} \\    
    \forall t, \sum_{j \neq i} b_{j,t} < \frac{|P|-|J|}{|J|}a_{i,t} + \frac{|P|}{|J|}
\end{align*}

These conditions mirror conditions (3) and (4) in the proof of Theorem 1. This is because for each litigant $i$, the worst-case scenario is that each other litigant $j$ coordinates to act as one coalition against $i$'s interests. And recall the necessity of these conditions (Theorem 2).

The following table shows the number of affirmatives $i$ must have when the other litigants have some total number of peremptories; and the number of peremptories $i$ must have when the other litigants have some total number of affirmatives.

\begin{center}
    \begin{tabular}{ |c||c|c|c|c| } 
     \hline
     Affirmatives & 0 & 1 & 2 & 3 \\ 
     \hline
     Peremptories & 0-1 & 0-2 & 1-3 & 2-3 \\ 
     \hline
    \end{tabular}
\end{center}

\hfill

We must consider three possible families of HAP methods to prove that no HAP method is quasi-proportional wrt each litigant.\\

Case 1: WLOG, L1 gets all 3 affirmatives. In this case, L2 must receive at least 2 peremptories before L1's third affirmative; and L3 must receive at least 2 peremptories before L1's third affirmative. But this totals to at least 7 actions, which is impossible, because $|P|=6$.\\

Case 2: WLOG, L1 gets 2 affirmatives and L2 gets 1 affirmative. In this case, L3 must receive at least 2 peremptories before the third affirmative is exercised; and L2 must receive at least 1 peremptory before L1's second affirmative is exercised. There are two subcases to consider:

\begin{itemize}
    \item L1's second affirmative is the last of the three affirmatives to be exercised. Then, both of L3's peremptories as well as L2's peremptory must come before L1's second affirmative. But the conditions would then be violated wrt L1, because 3 non-L1 peremptories would be exercised before L1 can exercise her second affirmative.

    \item L2's affirmative is the last of the three affirmatives to be exercised. Then, both of L3's peremptories must come before L2's affirmative. But the conditions would then be violated wrt L2, because 2 non-L2 peremptories would be exercised before L2 can exercise her affirmative.\\
\end{itemize}

Case 3: L1, L2, and L3 each get 1 affirmative. WLOG, suppose that L1's affirmative comes first, L2's second, and L3's third. Both L1 and L2 must receive their peremptory before L3's affirmative (because for each of them, L3's affirmative would be the second non-self affirmative). But the conditions would then be violated wrt L3, because 2 non-L3 peremptories would be exercised before L3 can exercise her affirmative.\\

Note that the presence of extra litigants beyond L1, L2, and L3 does not change this case analysis: at least one of L1, L2, and L3 will not be able to guarantee the proportionality of $J$ with respect to $P$ for all $S$.

\newpage

\section{Conclusion}

Given a setting in which two litigants must choose a number of jurors from a larger pool of potential jurors (the panel), I consider the problem of designing a jury selection method that allows each of the litigants to guarantee that the jury is proportional with respect to their preferences over the panel. I show that neither a pure system of peremptory strikes nor a pure system of affirmative selection provides such a guarantee, but identify a family of hybrid methods that combine the two that does. This family of methods is proportional even when litigants do not have opposing preferences, and have multi-tiered preferences. From a practical perspective, this family of methods should also be amenable to implementation due to its similarity to the existing status quo of peremptory strikes.

\newpage

\bibliography{biblio}
\nocite{*}

\end{document}